\documentstyle[amsfonts,aps,12pt]{revtex}

\begin{document}
\draft
\title{The Partition Function of a Spinor Gas }
\author{L. F. Lemmens}
\address{Departement Natuurkunde, Universiteit Antwerpen RUCA, Groenenborgerlaan 171,%
\\
B-2020 Antwerpen, Belgi\"{e}.}
\author{F. Brosens, J. T.\ Devreese.}
\address{Departement Natuurkunde, Universiteit Antwerpen UIA, Universiteitsplein 1,\\
B-2610 Antwerpen, Belgi\"{e}}
\date{October 21, 1999}
\maketitle

\begin{abstract}
For a spinor gas, i.e., a mixture of identical particles with several
internal degrees of freedom, we derive the partition function in terms of
the Feynman-Kac functionals of polarized components. As an example we study
a spin-1 Bose gas with the spins subjected to an external magnetic field and
confined by a parabolic potential. From the analysis of the free energy for
a finite number of particles, we find that the specific heat of this ideal
spinor gas as a function of temperature has two maxima: one is related to a
Schottky anomaly, due to the lifting of the spin degeneracy by the external
field, the other maximum is the signature of Bose-Einstein condensation.
\end{abstract}


\section{Introduction}

In this paper we extend our approach, introduced in \cite
{BDLPRE97a,BDLPRE98a} for a class of interacting polarized quantum systems,
to systems with internal degrees of freedom. After a review and a
generalization of our methods to the unpolarized case, their feasibility is
tested on a relatively simple confined model system of spin-1 bosons in an
external magnetic field. The partition function, the specific heat and the
susceptibility of this trapped spinor gas are evaluated for 100 and 1000
particles. Explicit analytical results in closed form were derived for 6
particles using symbolic algebra.

In the thermodynamic limit ($N\rightarrow \infty $ with a fixed density $%
\rho =N/V$), Ginibre \cite{GJMP65} provided a Feynman-Kac functional
description of the quantum statistical equilibrium for systems with internal
degrees of freedom, based on the grand canonical ensemble \cite{GGB71}. This
approach, applied to quantum systems \cite{LVVPRA93,VerVor} and quantum
plasma's \cite{CPRE96,APPRE96}, has recently been generalized by Cornu \cite
{CPRE98} to mixtures of identical particles in an external magnetic field. A
review of the methodology can be found in \cite{BMcondmat} with emphasis on
low-density Coulomb systems.

The approach which we present here is based on a different description \cite
{FeyBen72,Kubo}, although both Ginibre's and our approach have in common the
replacement of second quantization by a direct use of the permutation
symmetry in a path integral. But there are essential differences in the way
that the finite number of particles is incorporated in the theory.
Furthermore, we treat here an unpolarized mixture of identical particles.

Our approach avoids the thermodynamical limit and keeps the relations
between the density of states, the partition function for $N$ particles and
their generating function mathematically exact \cite{LBDSSC99}. These
statistical quantities are shown to be transforms of each other, i.e., they
are {\em not separately defined} on the basis of properties of an ensemble.
The partition function $Z_{N}\left( \beta \right) $ is the Laplace transform
of the distribution of states $\Omega _{N}\left( E\right) $ with $\beta $
adjoint to the energy $E$. The generating function $\Xi \left( \beta ,\gamma
\right) $ of the partition function is a Z-transform of $Z_{N}\left( \beta
\right) $ with the fugacity $\gamma $ adjoint to the number of particles.
For a harmonic model with interactions \cite{BDLPRE98a,LBDSSC99} we could
show, using the appropriate saddle-point methods reliable for the singular
structure of the generating function at large $N$, that these transforms
give the well known relation between the fugacity --and thereby the chemical
potential-- and the number of particles. Our derivation was largely inspired
by the inversion method used by Fowler and Darwin \cite{Kubo} to obtain the
relation between the temperature and the internal energy. The temperature as
a measure for the internal energy and the chemical potential as a
characterization of the number of particles are quantities obtained from the
theory, whereas in an ensemble-based approach (valid in the thermodynamic
limit), their definitions belong to the theory. A similar statistical
methodology has been recently proposed by Bormann et al. \cite{BorrPRA99} as
a new approach for systems with a finite number of particles, although their
method is only applicable for a partition function of a polarized system
without any particle-particle interaction.

Some models of harmonically confined systems with harmonic two-body
interactions can be solved exactly. They have been studied using other
techniques by many authors \cite{Sato,John,Angelucci,LBLBcm98,LBLBcm99}. The
classical version of the model was already studied by Newton \cite
{ChanOx95,N64}. The reason for the exact solvability is the reduction to a
center-of-mass evolution and an evolution of the other degrees of freedom
relative to the center of mass, the latter being independent of each other 
\cite{LBDPRE99}. Despite the conceptual simplicity of this approach, the
actual calculation turns out to be quite involved. Indeed, a general
two-body interaction requires a cumulant expansion of the exponential in the
Feynman-Kac functional with the implication that $n$-point correlation
functions have to be calculated for each higher order cumulant. These point
correlation functions have their own generating functions, which have to be
inverted separately. A loop, as is well known from a field theoretic
approach to the many-body problem, manifests itself in the irreducible part
of a $n$-point correlation function. In principle, these loops can be used
to simulate a many-body system \cite{LVVPRA93} or are used in a Mayer
cluster expansion \cite{CPRE96,Gruunp97} of the statistical quantities
relevant for stability studies of low-density Coulomb systems. Although this
prescription is general, we have applied it only to the first order cumulant 
\cite{TBLDSSC98}, in using the Jensen-Feynman variational principle to
optimize the free energy for realistic interaction potentials. The second
cumulant would already require a four-point correlation function.

Another extension of Ginibre's approach \cite{LVVPRA93,CPRE98}, considered
in this paper, concerns the spin degrees of freedom. We have to project the
full $N$-body propagator for distinguishable particles on the irreducible
representations of the permutation group. In the calculation of the
partition function of the model we only need the diagonal part with respect
to the spin degrees of freedom. This means that only those propagations have
to be considered where particles start and end in the same spin state after
some Euclidean time lapse. This simplification cannot be made in the
calculation of a $n$-point correlation function. In that case the evolution
of a spin state should be described by a continuous-time Markov process with
discrete states \cite{LPLA96}. These states serve as indices on the sample
paths in the Feynman-Kac functional. Also when charges are present, the
influence of the magnetic field on the charged particles can be taken into
account with the same methods \cite{FBDL99}.

The care given to the $N$-body aspects in our formulation originates from
the fact that a crossover from a density dependent behavior to a behavior
dependent on the number of particles is experimentally accessible. It has
become possible to produce assemblies of atoms whose internal energy is so
small, that quantum effects such as Bose-Einstein condensation or
Fermi-Dirac degeneracy become observable \cite
{AndSc95,DavPRL95,BraPRL95,BraPRL97}. The theoretical challenge posed by
these systems is their particular kind of confinement and the relatively low
number of atoms involved. This low number of atoms suggests that there is a
regime where the number of particles is more important than their density.
The study of such a crossover excludes the thermodynamical limit as an
investigation tool \cite{Ross}. The energy levels are primarily determined
by the trapping potential, rather than by the confining volume used in most
quantum-statistical studies. In the earliest experimental realizations of
Bose-Einstein condensation the system was restricted to a specific internal
degree of freedom; in present day studies this condition has been relaxed
leading to the so-called trapped spinor quantum gases \cite
{HSPRL96,MSSICKPRL,TiPRL98}, where an equilibrium over the internal degrees
of freedom can be reached.

In the present paper we illustrate the calculation techniques for a boson
gas with three internal degrees of freedom. A preliminary report on the
results of this investigation can be found in \cite{LBDPhysB00}. The paper
is organized as follows. In section II we derive the quantum mechanical
partition function of a mixture of identical particles, given only the total
number of particles. In order to achieve this goal, we go through a number
of steps: first we derive a partition function conditioned on a particular
state of the internal degrees of freedom and a given number of particles as
a Feynman-Kac functional \cite{Simon,ROEPS,SchWil81} that can be obtained
explicitly by path integral calculations, at least for the model under
consideration. The resulting propagator is projected on the symmetric or
antisymmetric irreducible representation of the permutation group according
to the boson or the fermion character of the particles. The cyclic
decomposition of the permutations imposes a constraint on the summations
over the cycle lengths. This difficulty is circumvented by introducing a
generating function for each conditioned partition function, which can be
inverted. The partition function of the mixture is then obtained as an
appropriate combination of the conditioned partition functions, from which
the free energy and the other thermodynamical quantities of the mixture
result. We describe these steps using an harmonic boson model, supplied with
an homogeneous external magnetic field, i.e., with a Zeeman splitting of the
spin degrees of freedom. In section III the free energy, the susceptibility
and the specific heat will be obtained with emphasis on the low temperature
properties and the thermal fluctuations of a boson gas. Finally, in section
IV we discuss the method and the obtained results, and we conclude the paper
with a brief summary of the work.

\section{Quantum statistics of mixtures}

In this section we provide the necessary background material to formulate a
quantum statistical theory of mixtures of a fixed number of $N$ identical
particles with internal degrees of freedom. Using the same approach as we
developed \cite{BDLPRE97a,BDLPRE98a} for particles without internal degrees
of freedom, we have to describe the state space and the evolution on that
state space using a process and a Feynman-Kac functional averaged over that
process. We also have to indicate how the projection on the irreducible
representations of the permutation group has to be done in order to ensure
the indistinguishability of those particles that are in the same state of
their internal degrees of freedom. Finally, we give an expression for the
partition function of the mixture.

\subsection{The propagator of the harmonically confined model}

The quantum model which we consider contains $N$ identical particles kept
together in a confining potential given by 
\begin{equation}
V_{1}=\frac{1}{2}\sum_{m=-s}^{s}\Omega _{m}^{2}\sum_{k=1}^{N_{m}}{\bf r}%
_{m,k}^{2},
\end{equation}
where ${\bf r}_{m,k}$ is the position vector of the $k^{\text{th}}$ particle
in state $m.$ (Natural units with $\hbar $ and the particle mass equal to
unity are used througout this paper.) The frequencies $\Omega _{m}$ are
related to the curvature of the parabolic confinement potential, which in
principle might be different for each internal degree of freedom. We assume
that there are $N_{m}$ particles that occupy the state $m.$ The total number
of particles is obtained from 
\begin{equation}
N=\sum_{m=-s}^{s}N_{m}.
\end{equation}
In the absence of interparticle interactions and for {\sl distinguishable}
particles, the propagator for the spatial degrees of freedom can be written
as a product 
\begin{equation}
K_{D}\left( \left\{ {\bf r}_{m,k}^{\prime \prime }\right\} ,\beta |\left\{ 
{\bf r}_{m,k}^{\prime }\right\} ,0\right)
=\prod_{m=-s}^{s}\prod_{k=1}^{N_{m}}\left. K\left( {\bf r}_{m,k}^{\prime
\prime },\beta |{\bf r}_{m,k}^{\prime },0\right) \right| _{\Omega _{m}},
\end{equation}
where $\left\{ {\bf r}_{m,k}^{\prime }\right\} $ represents a configuration
of all the particles in the internal state $m,$ and with 
\begin{equation}
\left. K\left( {\bf r}_{a},\beta |{\bf r}_{b},0\right) \right| _{w}=\left( 
\frac{w}{2\pi \sinh w\beta }\right) ^{\frac{3}{2}}\exp \left[ -\frac{w}{2}%
\frac{\left( {\bf r}_{a}^{2}+{\bf r}_{b}^{2}\right) \cosh w\beta -2{\bf r}%
_{a}\cdot {\bf r}_{b}}{\sinh w\beta }\right] .
\end{equation}

We assume that the internal degrees of freedom behave like a spin in an
external homogeneous magnetic field ${\bf B,}$ described by the Hamiltonian 
\begin{equation}
{\cal H}=-\mu {\bf B}\cdot {\bf S.}
\end{equation}
Introducing for each particle the states characterized by the value $m$ of
the spin component along the quantization axis 
\begin{equation}
S_{z}\left| s,m\right\rangle =m\left| s,m\right\rangle ,\quad
m=-s,-s+1,\ldots ,s,
\end{equation}
one finds that the Euclidean time evolution for the spin states of the
particle is given by: 
\begin{equation}
{\cal K}_{s}\left( m^{\prime \prime },\beta |m^{\prime },0\right)
=\left\langle s,m^{\prime \prime }\right| \exp -\beta {\cal H}\left|
s,m^{\prime }\right\rangle .
\end{equation}
This propagator can be written in the spectral representation by introducing
a unitary matrix $U$ that diagonalizes ${\cal H}$, 
\begin{equation}
U{\cal H}U^{-1}=-\mu \left| {\bf B}\right| S_{z}{\bf \ ,}
\end{equation}
as follows 
\begin{equation}
{\cal K}_{s}\left( m^{\prime \prime },\beta |m^{\prime },0\right) =\sum_{m%
{\bf =-}s}^{s}\left\langle s,m^{\prime \prime }\right| U^{-1}\left|
s,m\right\rangle e^{\beta \omega _{m}}\left\langle s,m\right| U\left|
s,m^{\prime }\right\rangle ,
\end{equation}
with $\omega _{m}=\mu m\left| {\bf B}\right| .$

Disregarding the statistics for the time being, the propagator for all the
(distinguishable) particles including the internal degrees of freedom thus
becomes: 
\begin{equation}
K_{D}\left( \left\{ {\bf r}_{m,k}^{\prime \prime }\right\} ,\beta |\left\{ 
{\bf r}_{m,k}^{\prime }\right\} ,0\right) =\prod_{m=-s}^{s}\left[
e^{N_{m}\beta \omega _{m}}\prod_{k=1}^{N_{m}}\left. K\left( {\bf r}%
_{m,k}^{\prime \prime },\beta |{\bf r}_{m,k}^{\prime },0\right) \right|
_{\Omega _{m}}\right] .
\end{equation}

The quantum evolution of the system is now defined as follows: the particles
are confined by a harmonic potential that may differ according to their spin
state $m$, and the state $m$ of the $j^{th}$ particle evolves as a spin-$m$
state in a magnetic field. This model is an idealization because interaction
between the spins of different particles is left out, spin-orbit coupling or
pair formation is not considered, and there is no spatial two-body
interaction. But in the assumption that these interactions can be included,
this simplified model still has to be studied as a zero order approximation.

\subsection{The permutation symmetry and the generating function}

From the propagator $K_{D}$ for the distinguishable particles, the
propagator $K_{mixt}$ of the mixture, taking into account the fermion or
boson statistics, can be obtained by using the antisymmetric or symmetric
projection as documented in \cite{FeyBen72,Kubo}: 
\begin{equation}
K_{mixt}\left( \left\{ {\bf r}_{m,k}^{\prime \prime }\right\} ,\beta
|\left\{ {\bf r}_{m,k}^{\prime }\right\} ,0\right)
=\prod_{m=-s}^{s}e^{N_{m}\beta \omega _{m}}{\Bbb K}_{I}\left( \left\{ {\bf r}%
_{m,k}^{\prime \prime }\right\} ,\beta |\left\{ {\bf r}_{m,k}^{\prime
}\right\} ,0\right) .
\end{equation}
The expression for the propagator ${\Bbb K}_{I}\left( \left\{ {\bf r}%
_{m,k}^{\prime \prime }\right\} ,\beta |\left\{ {\bf r}_{m,k}^{\prime
}\right\} ,0\right) $ of identical particles in the same spin state $m$ is 
\begin{equation}
{\Bbb K}_{I}\left( \left\{ {\bf r}_{m,k}^{\prime \prime }\right\} ,\beta
|\left\{ {\bf r}_{m,k}^{\prime }\right\} ,0\right) =\frac{1}{N_{m}!}%
\sum_{P}\xi ^{P}\prod_{k=1}^{N_{m}}\left. K\left( {\bf r}_{m,P\left(
k\right) }^{\prime \prime },\beta |{\bf r}_{m,k}^{\prime },0\right) \right|
_{\Omega _{m}},
\end{equation}
which represents a Feynman-Kac functional of the polarized components with
spin state $m$, including all their permutations $P$. Knowing the
propagator, the partition function for a given distribution $N_{-s},\ldots
,N_{s}$ can be expressed as a multiple integral: 
\begin{equation}
Z\left( \beta |N_{-s},\ldots ,N_{s}\right) =\left[ \prod_{m=-s}^{s}%
\prod_{k=1}^{N_{m}}\int d{\bf r}_{m,k}\right] \prod_{m=-s}^{s}e^{N_{m}\beta
\omega _{m}}{\Bbb K}_{I}\left( \left\{ {\bf r}_{m,k}\right\} ,\beta |\left\{ 
{\bf r}_{m,k}\right\} ,0\right) .
\end{equation}
The calculation for each spin degree of freedom then proceeds in complete
analogy with the polarized case \cite{BDLPRE97a}, leading to a partition
function of a mixture with a given composition of internal degrees of
freedom 
\begin{equation}
Z\left( \beta |N_{-s},\ldots ,N_{s}\right) =\prod_{m=-s}^{s}e^{N_{m}\beta
\omega _{m}}{\Bbb Z}\left( \beta |N_{m}\right) ,
\end{equation}
where ${\Bbb Z}\left( \beta |N_{m}\right) $ is the partition function of $%
N_{m}$ spin-polarized particles: 
\begin{equation}
{\Bbb Z}\left( \beta |N_{m}\right) =\left[ \prod_{k=1}^{N_{m}}\int d{\bf r}%
_{m,k}\right] {\Bbb K}_{I}\left( \left\{ {\bf r}_{m,k}\right\} ,\beta
|\left\{ {\bf r}_{m,k}\right\} ,0\right) .
\end{equation}

Of course, if the total number $N$ of particles is fixed, all combinations
of the $N_{-s},\ldots ,N_{s}$ (with $\sum_{m=-s}^{s}N_{m}=N$) are possible
and we obtain the following expression for the partition function of the
mixture: 
\begin{equation}
{\frak Z}_{mixt}\left( \beta |N\right)
=\sum_{N_{-s}=0}^{N}\sum_{N_{-s+1}=0}^{N}\ldots \sum_{N_{s}=0}^{N}C\left(
N_{-s},\ldots ,N_{s}\right) Z\left( \beta |N_{-s},\ldots ,N_{s}\right) ,
\end{equation}
\begin{equation}
C\left( N_{-s},\ldots ,N_{s}\right) =\frac{N!\Theta \left( N=\sum_{m{\bf =}%
-s}^{s}N_{m}\right) }{N_{-s}!N_{-s+1}!\ldots N_{s}!},
\end{equation}
where $C\left( N_{-s},\ldots ,N_{s}\right) $ is a multinomial coefficient in
which $\Theta \left( N=\sum_{m{\bf =}-s}^{s}N_{m}\right) $ expresses the
constraint on the summations, i.e., $\Theta \left( x\right) =1$ if the
logical variable $x$ is true, and $\Theta \left( x\right) =0$ if the logical
variable $x$ is false. Using the expression for the joint partition function 
${\frak Z}_{mixt}\left( \beta |N\right) $ in terms of the polarized
partition functions ${\Bbb Z}\left( \beta |N_{m}\right) $ one readily
obtains: 
\begin{equation}
{\frak Z}_{mixt}\left( \beta |N\right)
=\sum_{N_{-s}=0}^{N}\sum_{N_{-s+1}=0}^{N}\ldots \sum_{N_{s}=0}^{N}\left[
C\left( N_{-s},\ldots ,N_{s}\right) \prod_{m=-s}^{s}{\Bbb Z}\left( \beta
|N_{m}\right) e^{N_{m}\beta \omega _{m}}\right] .  \label{eq:Zmixture}
\end{equation}

This is the partition function that will be analyzed in the next section. We
consider there a gas of atoms with Bose-Einstein statistics confined in a
harmonic potential, without interparticle interactions, and with the ability
to adjust their internal degrees of freedom in order to attain
thermodynamical equilibrium. If the system is prepared in such a way that
the total $z$ component of all the spins can be fixed, then this constraint
can be taken into account using a technique developed in \cite{ElZEPR}. In
this case one has also to calculate expression (\ref{eq:Zmixture}) first.

\section{A spin-1 Bose gas}

In this section we illustrate the feasibility of the proposed approach on a
set of spin-1 bosons confined in a harmonic potential, based on the
partition function (\ref{eq:Zmixture}) of the mixture. For simplicity, the
many-body interaction, studied in \cite{BDLPRE97a,BDLPRE98a} to allow for
oscillations of the center of mass with a different frequency than that of
the oscillations of the internal degrees of freedom, is not taken into in
consideration here. This leads to a simpler albeit non trivial model. The
calculation techniques can almost entirely be based on those that have been
used for the polarized case \cite{BDLPRE97a,BDLPRE98a}.

The generating function $\Xi \left( \beta ,u\right) =\sum_{N_{m}=0}^{\infty }%
{\Bbb Z}\left( \beta |N_{m}\right) u^{N_{m}}$ of the partition function $%
{\Bbb Z}\left( \beta |N_{m}\right) $ of each component in the mixture is
known from these previous calculations, and given by: 
\begin{equation}
\Xi \left( \beta ,u\right) =\exp \left( \sum_{\ell =1}^{\infty }\frac{1}{%
\ell }\frac{b^{\frac{3}{2}\ell }u^{\ell }}{\left( 1-b^{\ell }\right) ^{3}}%
\right) \text{ with }b=e^{-\beta \Omega _{m}}.
\end{equation}

It should be noted that $u$ and $b$ depend on the parameters of the
spin-polarized subsystem. In the actual calculation we consider the same
confining frequencies $\Omega _{m}=w$ for each subsystem (not exploiting the
possibility of different confining potentials for different spin states).

There exist many ways to derive recursion relations between the partition
functions with a different number of particles \cite
{Grossman1,Grossman2,Kirsten,Ketterle,BorrFran}. A binomial expansion of the
generating function almost immediately leads to: 
\begin{equation}
{\Bbb Z}\left( \beta |N\right) =\frac{1}{N}\sum_{n=0}^{N-1}\left( \frac{%
b^{\left( N-n\right) /2}}{1-b^{N-n}}\right) ^{3}{\Bbb Z}\left( \beta
|n\right) .
\end{equation}
Combining the solutions for ${\Bbb Z}\left( \beta |N\right) $ from the
recursion relations, the partition function of the mixture becomes: 
\begin{equation}
{\frak Z}_{mixt}\left( \beta |N\right)
=\sum_{N_{-1}}\sum_{N_{0}=0}^{N-N_{-1}}\frac{N!\exp \left( \left(
2N_{-1}-N+N_{0}\right) \beta \omega _{s}\right) }{N_{-1}!N_{0}!\left(
N-N_{0}-N_{-1}\right) !}{\Bbb Z}\left( \beta |N_{-1}\right) {\Bbb Z}\left(
\beta |N_{0}\right) {\Bbb Z}\left( \beta |N-N_{0}-N_{-1}\right) .
\end{equation}
which is the key result of the present paper. Once this function is known
the thermodynamical quantities can be obtained in a straightforward manner.

\subsection{Specific heat and magnetic susceptibility}

The numerical representation of the results has to cope with very large
numbers. For accuracy reasons the following transformation proves to be
useful: 
\begin{equation}
{\frak Z}_{mixt}\left( \beta |N\right) =N!e^{N\beta \omega _{s}}b^{\frac{3}{2%
}N}S_{mixt},
\end{equation}
with 
\begin{equation}
S_{mixt}=\sum_{m=0}^{N}\sum_{n=0}^{N-m}\exp \left( -\beta \omega _{s}\left(
m+2n\right) \right) \frac{\chi _{N-m-n}\chi _{m}\chi _{n}}{\left(
N-m-n\right) !m!n!},
\end{equation}
where the scaling of the polarized partition function is given by: 
\begin{equation}
{\Bbb Z}\left( \beta |N\right) =b^{\frac{3}{2}N}\chi _{N},
\end{equation}
with the appropriate recursion relation: 
\begin{equation}
\chi _{N}=\frac{1}{N}\sum_{m=0}^{N-1}\frac{\chi _{m}}{\left(
1-b^{N-m}\right) ^{3}},\text{ with }\chi _{0}=1.
\end{equation}
Remembering the definition $\omega _{m}=\mu m\left| {\bf B}\right| ,$ it is
clear that $\omega _{s}$ denotes the maximal possible frequency due to the
magnetic field.

The expected number of particles $\left\langle N_{m}\left( \beta |N\right)
\right\rangle $ in each spin state $m$ then takes the following form
suitable for numerical computation: 
\begin{eqnarray}
\left\langle N_{1}\left( \beta |N\right) \right\rangle &=&\frac{1}{S_{mixt}}%
\sum_{m=0}^{N}\sum_{n=0}^{N-m}n\exp \left( -\beta \omega _{s}\left(
m+2n\right) \right) \frac{\chi _{N-m-n}\chi _{m}\chi _{n}}{\left(
N-m-n\right) !m!n!}, \\
\left\langle N_{0}\left( \beta |N\right) \right\rangle &=&\frac{1}{S_{mixt}}%
\sum_{m=0}^{N}\sum_{n=0}^{N-m}m\exp \left( -\beta \omega _{s}\left(
m+2n\right) \right) \frac{\chi _{N-m-n}\chi _{m}\chi _{n}}{\left(
N-m-n\right) !m!n!}, \\
\left\langle N_{-1}\left( \beta |N\right) \right\rangle &=&N-\left\langle
N_{0}\left( \beta |N\right) \right\rangle -\left\langle N_{1}\left( \beta
|N\right) \right\rangle .
\end{eqnarray}
The same parameterization can be used to obtain the internal energy of the
mixture 
\begin{equation}
U_{mixt}\left( \beta |N,\omega _{s}\right) =-\frac{d}{d\beta }\ln \left( 
{\frak Z}_{mixt}\left( \beta |N\right) \right) ,
\end{equation}
leading to 
\begin{eqnarray}
U_{mixt}\left( \beta |N,\omega _{s}\right) =N\left( \frac{3}{2}w-\omega
_{s}\right) +\frac{1}{S_{mixt}}\sum_{m=0}^{N}\sum_{n=0}^{N-m} &&e^{-\beta
\omega _{s}\left( m+2n\right) }\frac{\chi _{N-m-n}}{\left( N-m-n\right) !}%
\frac{\chi _{m}}{m!}\frac{\chi _{n}}{n!}  \nonumber \\
&&\times \left( U_{N-m-n}+U_{m}+U_{n}+\omega _{s}\left( m+2n\right) \right)
\end{eqnarray}
(with $U_{m}$ the internal energy of the particles in spin state $m).$ The
specific heat 
\begin{equation}
C\left( \beta |N,\omega _{s}\right) =\frac{d}{dT}U_{mixt}\left( \beta
|N,\omega _{s}\right)
\end{equation}
and the magnetic susceptibility $\frac{d}{dB}U_{mixt}\left( \beta |N\right)
, $ proportional to $\frac{dU\left( \beta |N,\omega _{s}\right) }{d\omega
_{s}} $, are easily obtained from this expression.

In fig. 1 we show the specific heat, the magnetic excess specific heat (i.e.
the contribution to the specific heat due to the magnetic field) and the
magnetic susceptibility for 100 bosons. For 1000 bosons the same quantities
are shown in fig 2. The temperature is expressed in units of $T_{c}=\frac{w}{%
k}\left( \frac{N}{\zeta \left( 3\right) }\right) ^{1/3}$, where $w$ is the
frequency parameter of the confinement, $k$ is the Boltzmann constant and $%
\zeta \left( 3\right) =\allowbreak 1.\,\allowbreak 202\,056\,903$ is a
Riemann zeta-function. The frequency parameter for the internal degrees of
freedom is expressed in units of $w$: 
\[
\omega _{s}=w_{s}w. 
\]
Using these units the expression for $b$ becomes $b=\exp \left( -\frac{1}{t}%
\left( \frac{\zeta \left( 3\right) }{N}\right) ^{1/3}\right) $ with $t=\frac{%
T}{T_{c}}$ and $\exp \left( -\beta \omega _{s}\right) =b^{w_{s}}.$ Because
there is a substantial dependence on the magnetic field strength, the
influence of the magnetic field can be illustrated by the redistribution of
the particles over the internal degrees of freedom, giving rise to the
magnetic excess specific heat. The magnetic susceptibility clearly
illustrates the dependence of the internal energy on the magnetic field
strength.

\subsection{The Schottky anomaly}

In the specific heat plotted in fig. 1 and fig. 2 we have identified the low
temperature maxima as Schottky anomalies. This contribution to the specific
heat is attributed to the lifting of the degeneracy of the levels of the
spin degrees of freedom, due to the magnetic field. The effect occurs
irrespective of the boson statistics, as will be illustrated below.

For up to 6 particles we calculated all polarized partition functions by
symbolic algebra. This allows to study the free energy and the derived
quantities exactly. Because the same calculation can be performed using
Maxwell-Boltzmann statistics, we may isolate the effect of the Bose-Einstein
statistics on the Schottky anomaly.

For numerically exact calculations it is useful to factorize out the
dominant behavior 
\begin{equation}
{\Bbb Z}\left( \beta |N\right) =z_{N}\left( \beta \right) \frac{b^{\frac{3N}{%
2}}}{\prod_{n=1}^{N}\left( 1-b^{n}\right) ^{3}}.
\end{equation}
The recursion relation for the resulting polynomials $z_{N}\left( \beta
\right) $ becomes then: 
\begin{equation}
z_{N}\left( \beta \right) =\frac{1}{N}\sum_{n=0}^{N-1}z_{n}\left( \beta
\right) \frac{\prod_{j=n+1}^{N}\left( 1-b^{j}\right) ^{3}}{\left(
1-b^{N-n}\right) ^{3}}.  \label{polynomials}
\end{equation}

The polynomials $z_{N}\left( \beta \right) $ for $N=1,\ldots ,6$ were
obtained by integer arithmetic and the results of this calculation are
listed in table 1. For low temperatures $\left( kT\ll w/2\right) $ the
specific heat is plotted in fig. 3 as a function of the temperature and of
the magnetic field. The contribution associated with the lifting of the
degeneracy and the evolution to a single polarized state with increasing
magnetic field clearly manifests itself. We have also studied the specific
heat of 6 {\em distinguishable} particles with spin degrees of freedom. The
result is shown in fig. 4, and it also exhibits a maximum due to the
Schottky anomaly. Comparing the specific heat of distinguishable particles
with that of bosons, it should be noted that the Bose-Einstein statistics
only weakly influences the magnitude of Schottky anomaly. But its relative
contribution is more pronounced for Bose-Einstein statistics than it is for
Maxwell-Boltzmann statistics because the boson character suppresses the
energy fluctuations at low temperature leading to a smaller specific heat.

\section{Discussion and conclusion}

First we review our method, then we comment on our results for the spin-1
model and finally we will conclude with a brief indication of the potential
use of the method.

A first striking difference between the method worked out here and more
standard approaches to the many-body problem is that we incorporate the
statistics using directly the representation theory of the symmetric group
instead of the more common second quantization. As we indicated before,
Ginibre's approach is based on the grand canonical ensemble, whereas we
based our method for particle-conserving systems on the distribution of
states, which we transformed to the partition function and its generating
function. The temperature and the chemical potential obtain their standard
meaning from the saddle-point method used to invert the transform. The
numerical analysis for the spin-polarized case published earlier \cite
{BDLPRE98a,LBDSSC99} indicates that when there are sufficiently many
particles, corrections to the saddle point inversion become negligible. We
believe that this remains the case for unpolarized systems but we did not
prove this yet.

When internal degrees of freedom are involved, we argued that a Feynman-Kac
functional averaged over an indexed Brownian motion describes the evolution
of the system. The projection on the appropriate irreducible representation
of the permutation group leads to a multinomial combination of the spin
states in the partition function. The index process to distinguish between
different spin states is not used explicitly, because only the diagonal part
of the spectral representation of the propagator of that process is needed
for the thermodynamics. The fact that we can construct a process for the
evolution of the internal degrees of freedom is nevertheless important, not
only for a better understanding of the behavior of the system at long
Euclidean times (low temperature) but also for considering correlations
between components in different internal states. We indicated briefly that
two-body interactions can be taken into account by a cumulant expansion of
the exponential in the Feynman-Kac functional. This implies that one-point,
two-point and $n$-point correlation functions have to be calculated. Other
expansions, like e.g. the Mayer cluster expansion, can presumably also be
carried out in our formalism, but this point deserves further investigation.

In order to demonstrate the feasibility of the approach, we gave an example
based on an exactly soluble harmonically confined spin-1 boson gas in a
magnetic field and calculated the internal energy, the magnetic
susceptibility and the specific heat for the system in equilibrium.
Furthermore, although we illustrated the technique without two-body
interactions, it should be noted that the variational method that we have
applied to the spin-polarized case can also be used to study mixtures of
identical particles with different degrees of freedom.

The example itself, a noninteracting boson gas, exhibits Bose-Einstein
condensation characterized by one of the maxima in the specific heat as a
function of temperature. A second maximum is due to the Schottky
contribution to the specific heat. It should be noted that the Schottky
anomaly indicates that the $z$-component of the total spin, which is zero in
the ground state, acquires values different from zero. The fact that this
occurs at a lower temperature than the BEC for an ideal system may of
experimental importance.

We conclude that the methods which we developed for polarized systems of
identical particles can also be extended to non-polarized systems, as shown
in the present paper. The basic tools are a Feynman-Kac functional with an
adapted process, the projection on the symmetric or anti-symmetric
representations, and the path integral evaluation of the averages.

\section{Acknowledgments}

This work is performed within the framework of the FWO\ projects No.
1.5.729.94, 1.5.545.98, G.0287.95, G.0071.98 and WO.073.94N
(Wetenschappelijke Onderzoeksgemeenschap over ``Laagdimensionele
systemen''), the ``Interuniversitaire Attractiepolen -- Belgische Staat,
Diensten van de Eerste Minister -- Wetenschappelijke, Technische en
Culturele Aangelegenheden'', and in the framework of the BOF\ NOI 1997
projects of the Universiteit Antwerpen.

\begin{center}
{\bf Table}

\begin{table}[tbp] \centering  %
%
\begin{tabular}{|l|l|}
$N$ & Polynomials $z_{N}\left( \beta \right) $ in $b=e^{-\beta w}$ [see Eq. (%
\ref{polynomials})] \\ \hline
$0$ & $1$ \\ 
$1$ & $1$ \\ 
$2$ & $1+3b^{2}$ \\ 
$3$ & $1+3b^{2}+7b^{3}+6b^{4}+6b^{5}+10b^{6}+3b^{7}$ \\ 
$4$ & $
\begin{array}[t]{l}
1+3b^{2}+7b^{3}+\allowbreak
18b^{4}+21b^{5}+47b^{6}+57b^{7}+87b^{8}+80b^{9}+\allowbreak 87b^{10}+63b^{11}
\\ 
+62b^{12}+27b^{13}+15b^{14}+b^{15}
\end{array}
$ \\ 
$5$ & $
\begin{array}[t]{l}
1+3b^{2}+7b^{3}+18b^{4}+39b^{5}+74b^{6}+138b^{7}+\allowbreak
252b^{8}+396b^{9}+600b^{10}+828b^{11} \\ 
+1087b^{12}+1314b^{13}+1503b^{14}+1585b^{15}+1560\allowbreak
b^{16}+1416b^{17}+1197b^{18}+921b^{19} \\ 
+669\allowbreak b^{20}+413b^{21}+231b^{22}+105b^{23}+37b^{24}+6b^{25}
\end{array}
$ \\ 
$6$ & $
\begin{array}[t]{l}
1+3\allowbreak
b^{2}+7b^{3}+18b^{4}+39b^{5}+99b^{6}+180b^{7}+390b^{8}+715b^{9}+1323b^{10}+2214b^{11}
\\ 
+3713b^{12}+5664b^{13}+8589b^{14}+\allowbreak
12\,129b^{15}+16\,683b^{16}+21\,606b^{17}+\allowbreak
27\,312b^{18}+32\,442b^{19} \\ 
+37\,569b^{20}+41\,147b^{21}+\allowbreak
43\,674b^{22}+43\,857b^{23}+42\,756b^{24}+39\,219\allowbreak
b^{25}+34\,854b^{26} \\ 
+29\,120b^{27}+23\,436b^{28}+\allowbreak
17\,565b^{29}+12\,733b^{30}+8388b^{31}+5304b^{32}+2994b^{33}+\allowbreak
1590b^{34} \\ 
+687b^{35}+290b^{36}+75b^{37}+15b^{38}
\end{array}
$%
\end{tabular}
\caption{In this table we list the polynomials $z_{N}\left( \beta \right) $ for
$N=1,\ldots ,6.$ The coefficients are obtained using integer arithmetic and
symbolic algebra.\label{key}}%
\end{table}  %
%

{\bf Figure captions}
\end{center}

\begin{description}
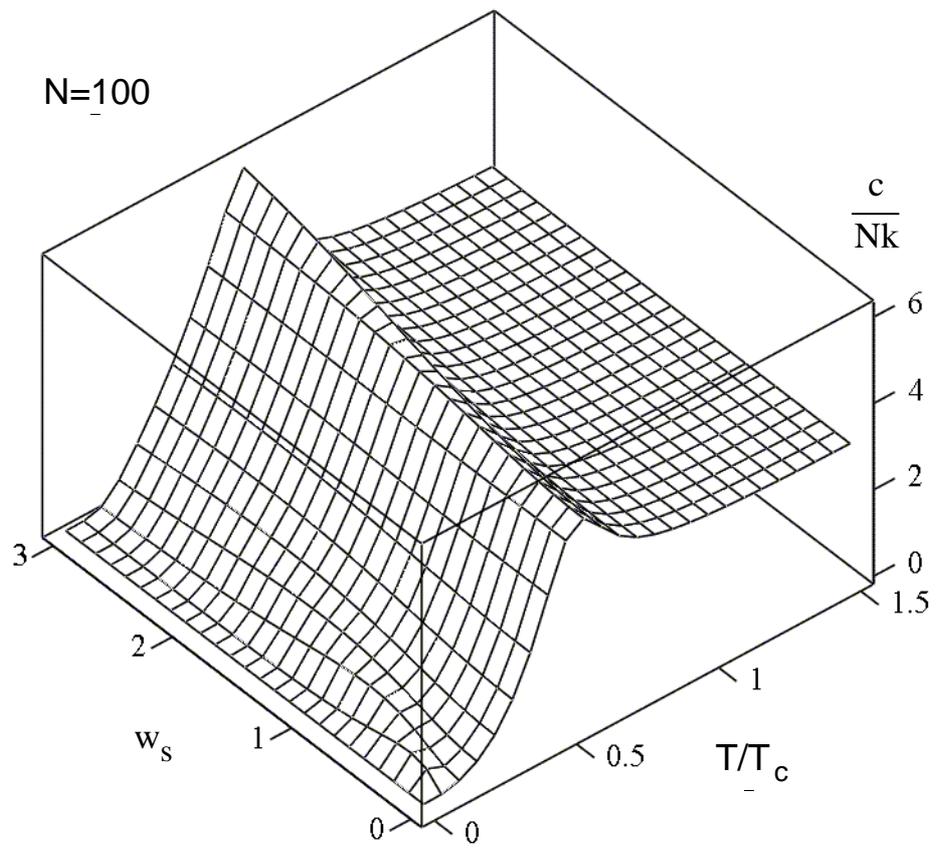
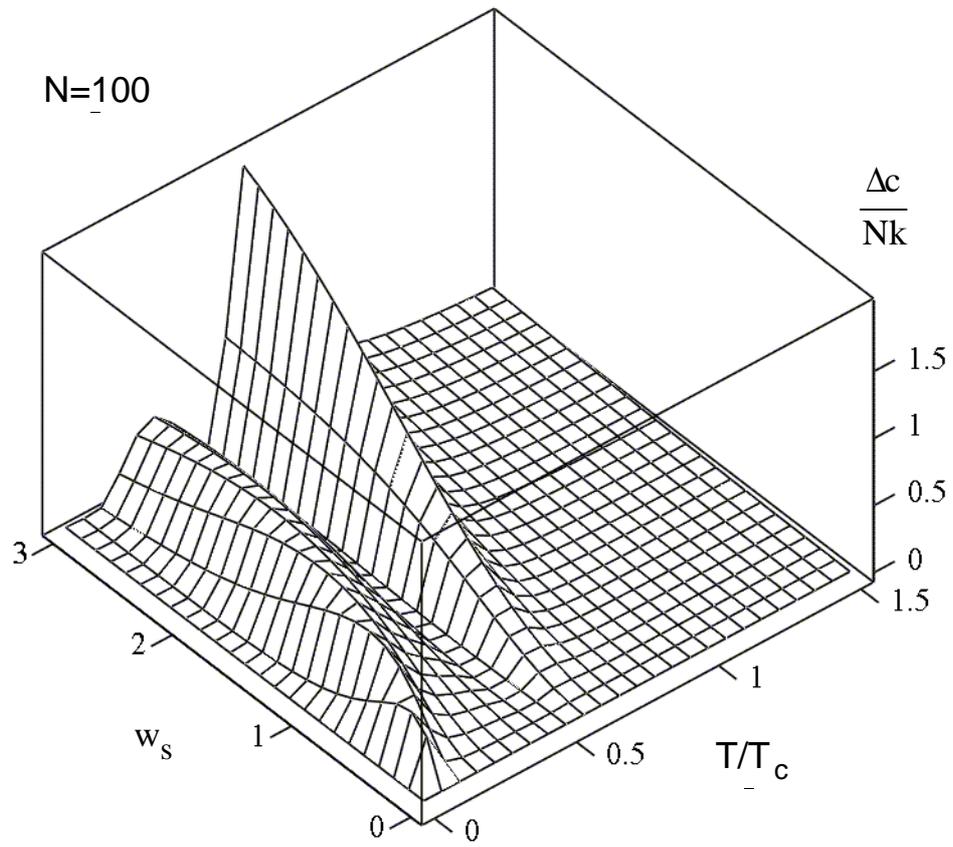
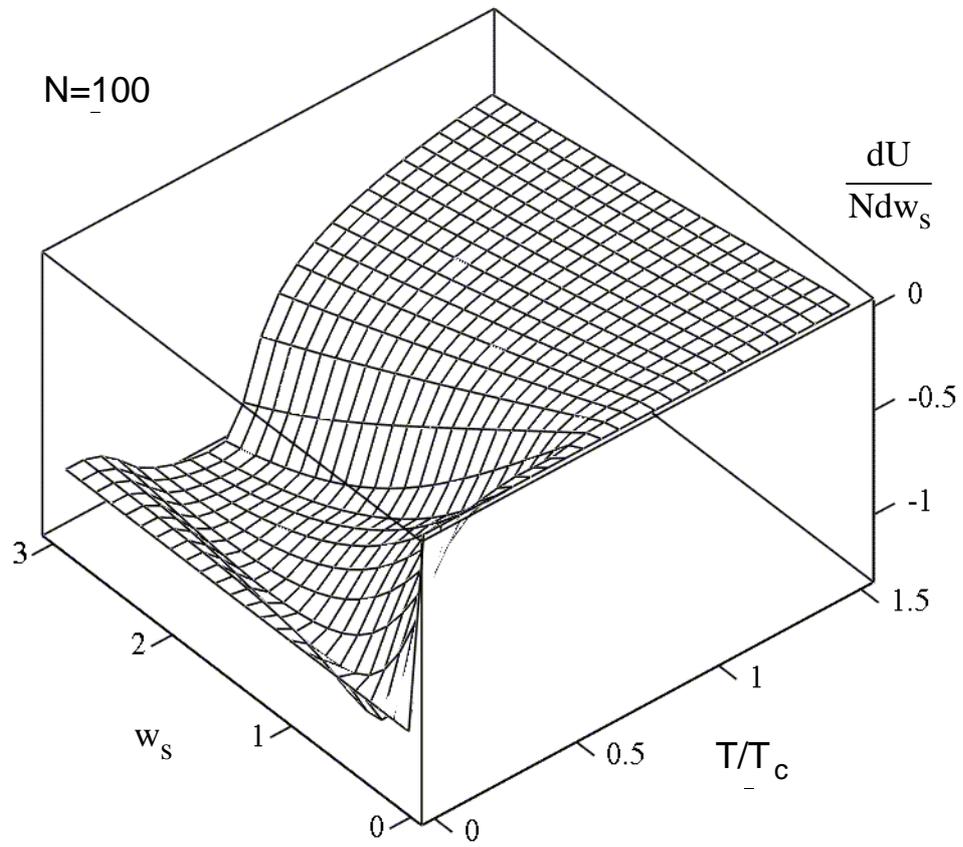
\item[Fig. 1:]  For 100 bosons the specific heat per particle (a), the
magnetic excess specific heat per particle (b), and the magnetic
susceptibility (c) are shown as a function of the temperature and the
magnetic field.

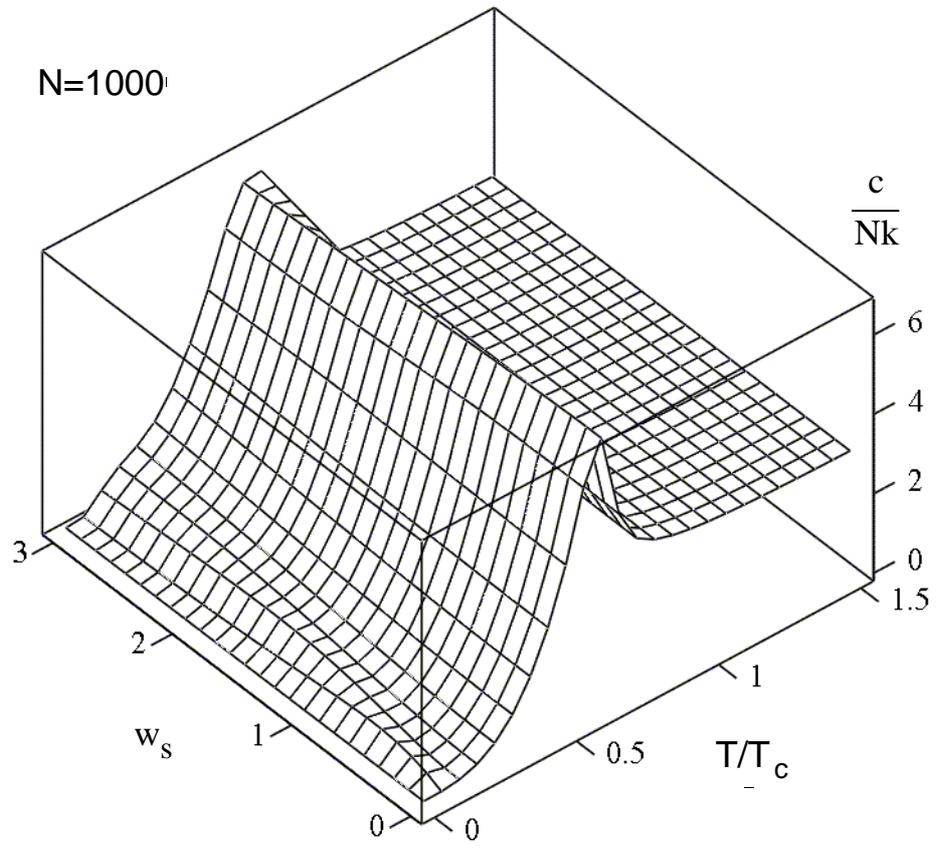
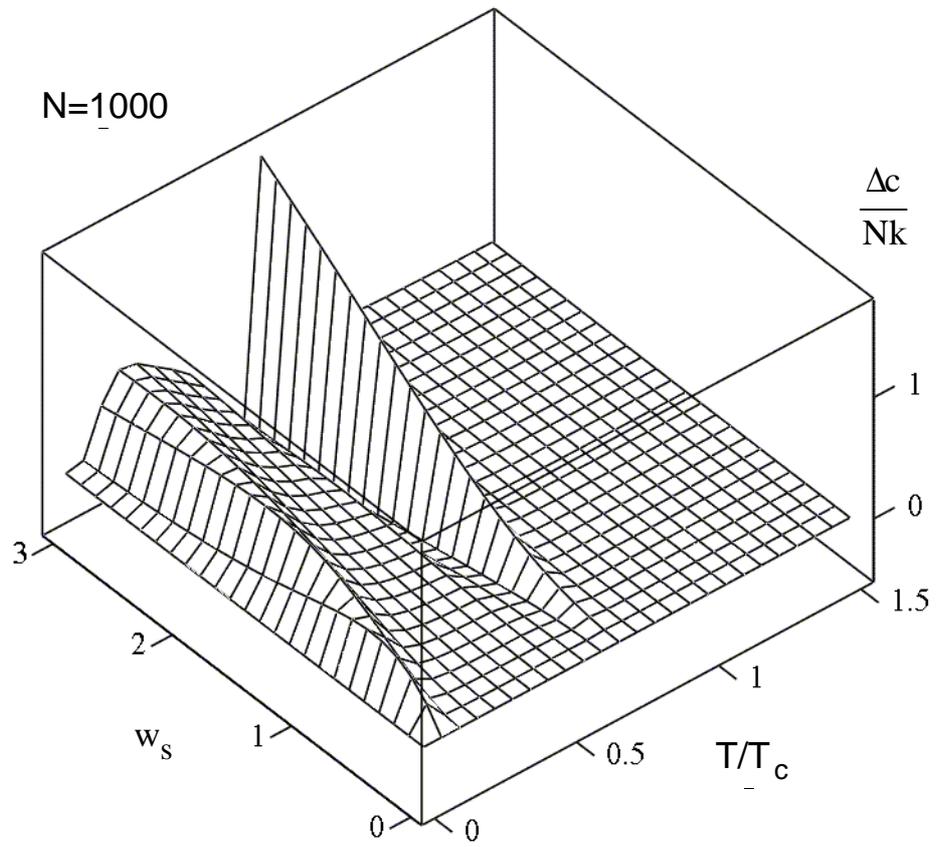
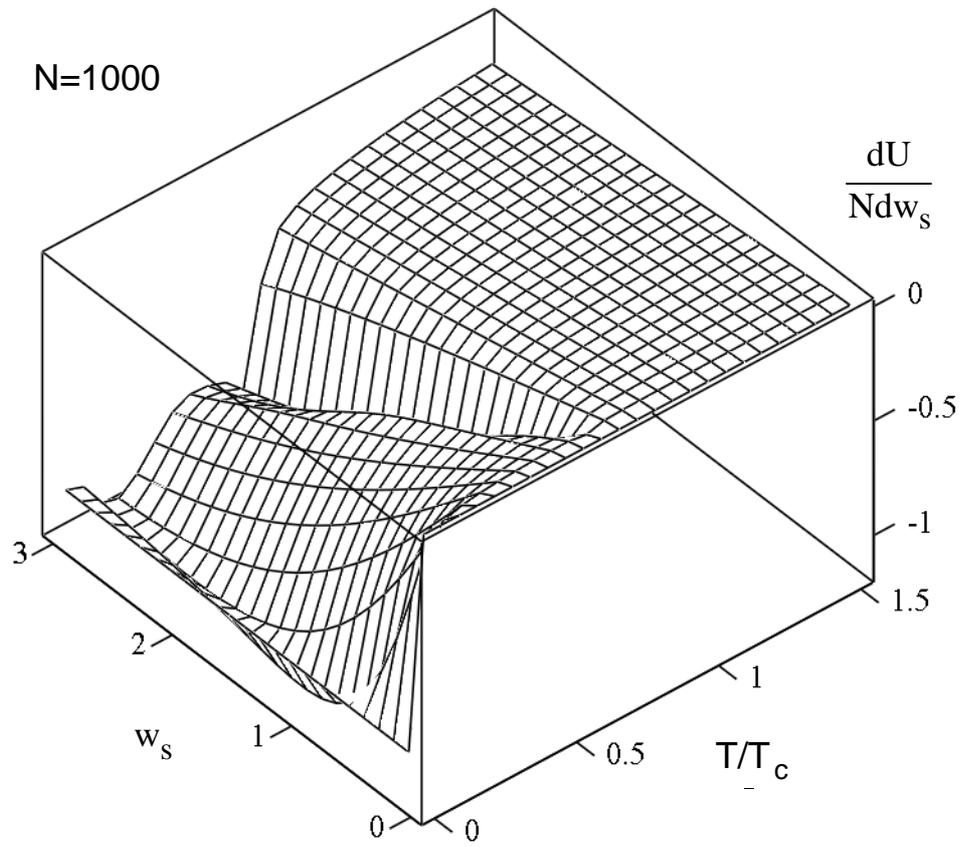
\item[Fig. 2:]  For 1000 bosons the specific heat per particle (a), the
magnetic excess specific heat per particle (b), and the magnetic
susceptibility (c) are shown on the same scale as in fig 1.

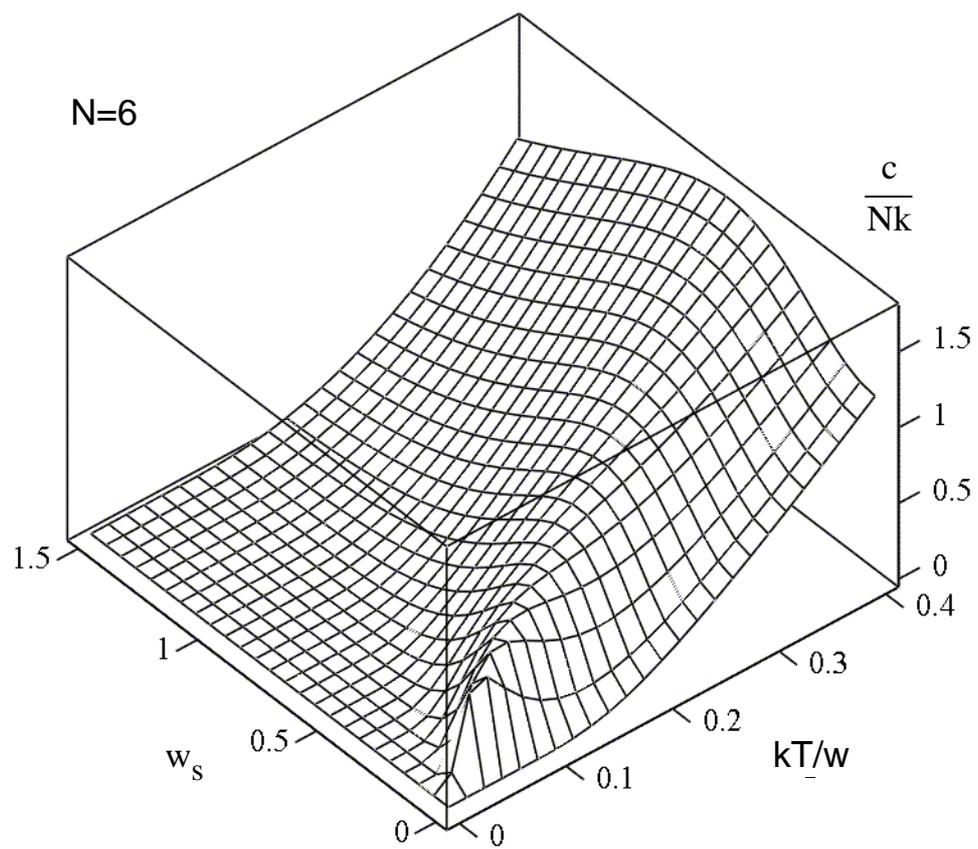
\item[Fig. 3:]  The specific heat per particle is shown for 6 bosons
distributed over 3 spin states in thermal equilibrium. The maximum in the $%
\left( C,w_{s}\right) $-plane with fixed $T$ is identified as the Schottky
anomaly. This is an entropic effect due to the lifting of the degeneracy of
the spin degrees of freedom by an external field.

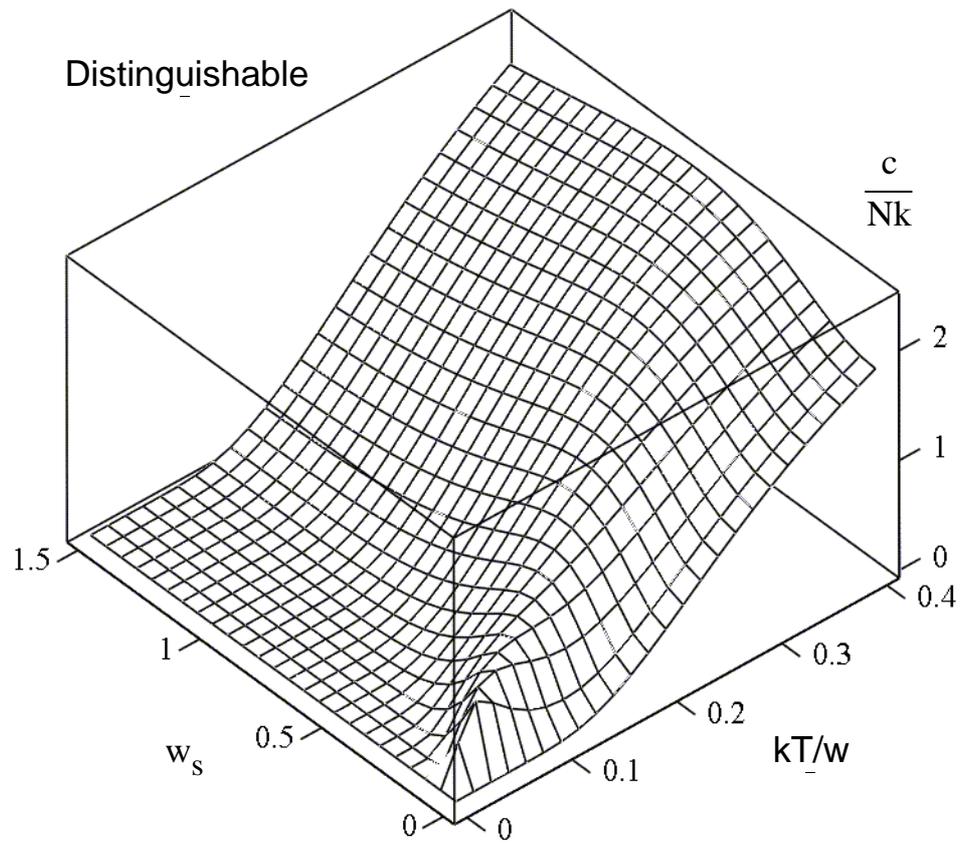
\item[Fig. 4:]  The specific heat per particle for 6 distinguishable
particles is shown with Maxwell-Boltzmann statistics. Note the presence of
the Schottky anomaly. This figure should be compared with figure 3 where the
same quantity is shown taking the Bose Einstein statistics into account.
\end{description}

\end{document}